# Semiempirical Modeling of Reset Transitions in Unipolar Resistive-Switching Based Memristors

*Rodrigo PICOS[1], Juan Bautista ROLDAN[2], Mohamed Moner AL CHAWA[1],*
*Pedro GARCIA-FERNANDEZ[2], Francisco JIMENEZ-MOLINOS[2], Eugeni GARCIA-MORENO[1]*

[1] Physics Dept., Universitat de les Illes Balears, Palma de Mallorca, 07122, Spain
[2] Departamento de Electrónica y Tecnología de los Computadores. Universidad de Granada, 18071, Spain

rodrigo.picos@uib.es,  jroldan@ugr.es

**Abstract.** *We have measured the transition process from the high to low resistivity states, i.e., the reset process of resistive switching based memristors based on Ni/HfO$_2$/Si-n+ structures, and have also developed an analytical model for their electrical characteristics. When the characteristic curves are plotted in the current-voltage (I-V) domain a high variability is observed. In spite of that, when the same curves are plotted in the charge-flux domain (Q-ϕ), they can be described by a simple model containing only three parameters: the charge ($Q_{rst}$) and the flux ($\phi_{rst}$) at the reset point, and an exponent, n, relating the charge and the flux before the reset transition. The three parameters can be easily extracted from the Q-ϕ plots. There is a strong correlation between these three parameters, the origin of which is still under study.*

## 1. Introduction

Resistive switching memories (RRAMs) are one of the most promising alternatives among emerging non-volatile memory technologies [1-3]. Resistive switching (RS) phenomena have been reported as early as the 1960s. These devices show very interesting features such as low program/erase currents, fast speed, endurance, viability for 3D memory stacks and CMOS technology compatibility [1–3].

RRAMs belong to a wider group of electron devices known as memristors [4]. These recently fabricated devices [5], whose theoretical features were predicted a long time ago [6], have shown many different possibilities ranging from non-volatile memory cells to neuromorphic circuit applications [1]. There are many technologies available to fabricate memristors [1], [2]. It has been shown that the physical background behind the device characteristics can be linked to different operation principles. Memristors based on electrochemical metallization phenomena show different features in comparison to devices controlled by thermochemical or valence change mechanisms. Therefore, many characterization, simulation and modeling studies are needed to understand the operation of these devices and to build the circuit simulation infrastructure needed to design memristor-based applications. Different modeling [5], [7–10] and simulation [11–16] studies have been published so far devoted to different flavors of memristive devices although there is still a long way to go in this field.

In this paper we have focused on the modeling of memristive devices to characterize some of their main parameters. To do so, we have used devices fabricated and measured at the IMB-CNM (CSIC) in Barcelona [17]. The devices were based on Ni/HfO$_2$/Si-n$^+$ structures fabricated on (100) n-type CZ silicon wafers. The resistive switching mechanisms of these devices were characterized and studied previously [11], [17].

The manuscript is organized in the following manner: In Sec. 2 we characterized the experimental measurements to use them for modeling purposes; in Sec. 3 we introduce our model and, finally, we draw the main conclusions in Sec. 4.

## 2. Memristor Experimental Analysis for Modeling Purposes

Several hundreds of resistive switching (RS) cycles were performed and measured on the devices described above after a forming process [17]. Some of the reset (process that leads the device from a low resistance state to a high resistance state) I-V curves have been plotted in Fig. 1 for a single device. It must be noted that, as reported previously [17], there is a great dispersion in the shapes of the curves and in the reset current and voltages (see the definition of these magnitudes in [18]). Although it is commonly accepted and used the device characterization in terms of I-V curves, we have chosen a different approach following the lines highlighted by L. Chua and others in the past [4], [9]. In this respect, instead of using an I-V domain we will work in a domain characterized by the charge (Q) and flux (ϕ) magnitudes. The flux is defined as the time integral of the voltage, while the charge is obtained as the time integral of the current [4]. Thus, Fig. 2 plots again the data of Fig. 1, but in the Q-ϕ domain.

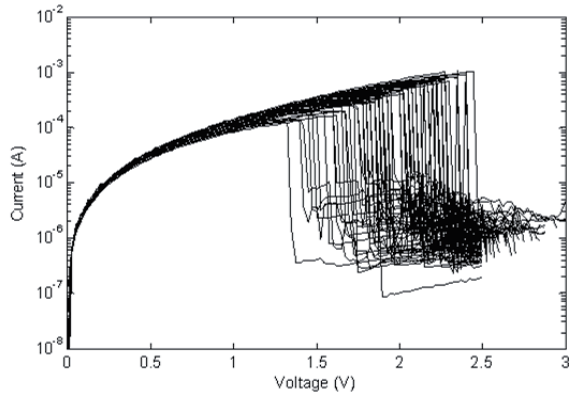

**Fig. 1.** I-V characteristics of reset transitions corresponding to 100 RS consecutive cycles on a single device.

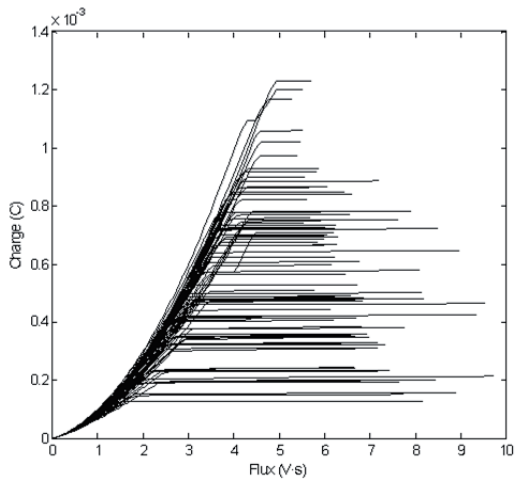

**Fig. 2.** Q-ϕ characteristics for the same reset cycles shown in Fig 1.

It is worth noticing that, as expected, after the reset voltage is achieved, the charge remains practically constant since the device current is reduced in several orders of magnitude and there is no contribution to the time integral of the charge. As reported in [11–16], the characteristic conductive filaments responsible for the RS operation in conductive bridge cells are broken when the reset process takes place. We take profit from this fact to extract the reset voltage ($V_{rst}$) and the reset current ($I_{rst}$). In fact, we first extract the reset flux ($\phi_{rst}$) and the corresponding reset charge ($Q_{rst}$); afterwards we recover the corresponding reset voltage and current values. To extract $Q_{rst}$ and $\phi_{rst}$ values, we fit two lines in the Q-ϕ plot (Fig. 3): one to the region where the charge remains constant and the other to the upper part of the monotonously increasing curve prior to the plateau. The desired reset point is given by the intersect point of the crossing lines, as depicted in Fig. 3a and 3b. The first figure shows the application of the method to a curve with a single current step, while Fig. 3b shows the application of this algorithm to a curve corresponding to a cycle with several current steps (an in-depth study on the physics behind these multisteps reset curves can be found in [16]). From these values ($\phi_{rst}$ and $Q_{rst}$, respectively), it is straightforward to calculate $V_{rst}$ and $I_{rst}$, since the flux and charge are monotonically increasing functions.

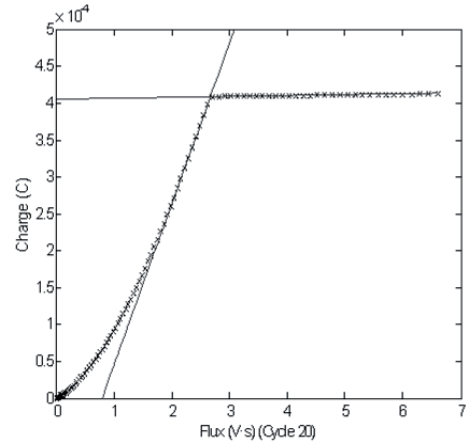

(a)

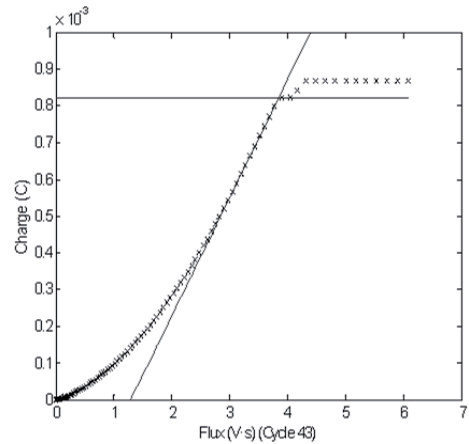

(b)

**Fig. 3.** Q versus ϕ for two different reset curves: (a) shows a single plain transition, while (b) shows a multiple transition. The fitting lines are shown as continuous, while the crosses are the transformed measured data. The reset point is assumed to be located at the intersection point between the lines.

| | $\phi_{rst}$ (Vs) | $Q_{rst}$ (mC) | $V_{rst}$ (V) | $I_{rst}$ (mA) |
|---|---|---|---|---|
| Average | 3.28 | 0.56 | 1.98 | 0.167 |
| Standard dev. | 0.76 | 0.25 | 0.26 | 0.088 |

**Tab. 1.** Mean values and standard deviations of the obtained reset values for the flux, charge, voltage and current.

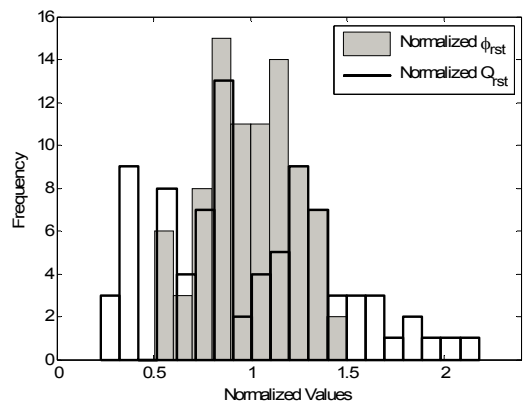

**Fig. 4.** Histogram of $Q_{rst}$ and $\phi_{rst}$ values, the data plotted have been previously normalized to the mean values shown in Tab. 1.

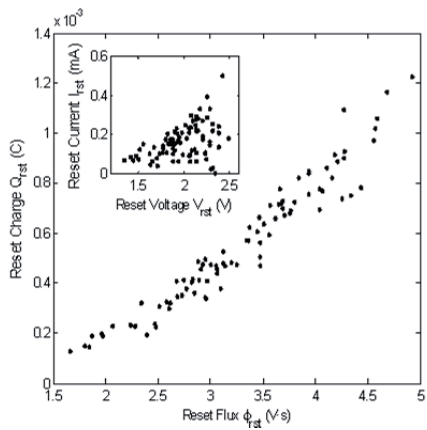

**Fig. 5.** $Q_{rst}$ vs $\phi_{rst}$ for each reset transition. The inset shows $I_{rst}$ versus $V_{rst}$. The correlation is clearly much better in the first case.

The mean values and standard deviations for all these calculated parameters are provided in Tab. 1. Figure 4 shows the statistical distribution of $\phi_{rst}$ and $Q_{rst}$. It has to be noted that only 100 cycles is not enough to distinguish whether the distribution is Gaussian or not; nevertheless, since this is not the key issue of this paper, we have not deepen on the statistical properties of the devices.

Figure 5 plots $Q_{rst}$ versus $\phi_{rst}$ and $I_{rst}$ versus $V_{rst}$ (in the inset). It is worth calling the reader attention to the fact that the correlation between them is much higher in the flux-charge variables than in the current-voltage pairs. This result is coherent with the fact that the Q-ϕ domain is the natural space for memristor modeling [4].

## 3. Semiempirical Model

Once the $\phi_{rst}$, $Q_{rst}$, $V_{rst}$ and $I_{rst}$ are calculated, we can normalize the measurements, scaling the curves in the Q-ϕ domain in order to obtain $Q_{rst} = 1$ and $\phi_{rst} = 1$. Figure 6 shows the results of the normalized charge versus the normalized flux. It is apparent that the behavior of the whole set of curves is very similar. In this respect, taking into consideration the work of Chua [4] and Shin [9] outlining a work plan for modeling in the Q-ϕ domain, we propose a non-linear relation between the charge and the flux in the following way (we assume a voltage-controlled memductor whose constitutive relationship is the following equation [9]):

$$Q = Q_{rst} \cdot \min\left(1, \left(\frac{\phi}{\phi_{rst}}\right)^n\right). \quad (1)$$

Notice that we have not taken into account any residual charge after the reset transition. This simple and explicit model fits the experimental data in a reasonable manner with only three parameters ($Q_{rst}$, $\phi_{rst}$, $n$). Some examples of this are shown in Fig. 7 and 8, where three random curves are depicted. The model fits experimental values fairly well, mostly in the Q-ϕ domain. The mean values of the fitting parameters are provided in Tab. 2. The values of $\phi_{rst}$ and $Q_{rst}$ are obviously the same than those given in Tab. 1, but are repeated here for completeness.

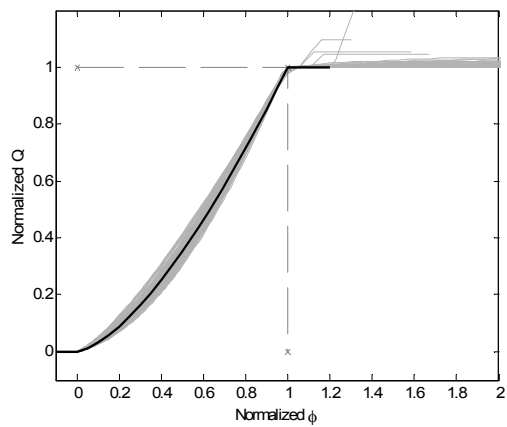

**Fig. 6.** Normalized $Q$ vs normalized $\phi$ for each reset transition. The normalization is performed by scaling each curve in order to fit the reset values to one. The darkest line corresponds to the model proposed in (1). In the case of I-V curves with several current steps, there can be seen Q values above one; in these cases the normalizing value was connected with the first reset event.

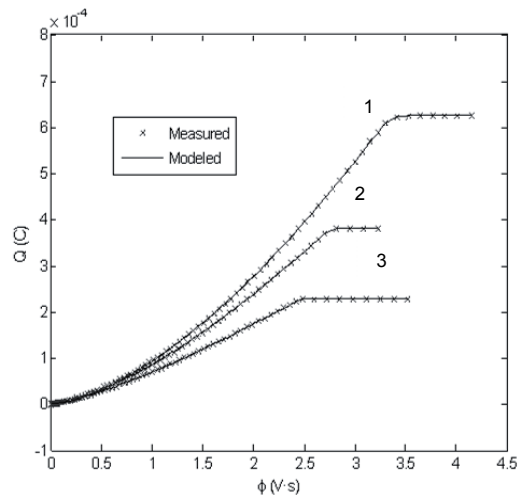

**Fig. 7.** Experimental (crosses) and modeled (lines) behavior of three different transitions in the Q-ϕ domain. The numbers refer to the corresponding curves in Fig. 8.

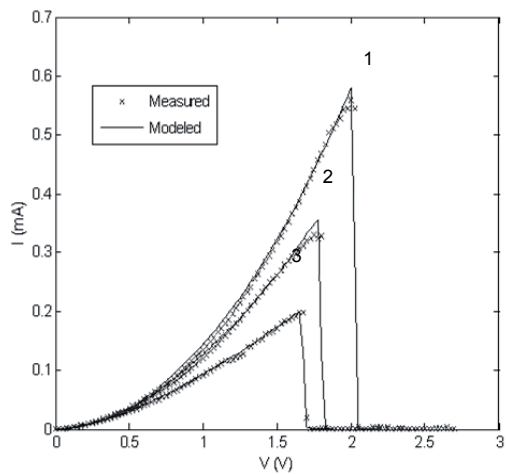

**Fig. 8.** Experimental (crosses) and modeled (lines) behavior of three different transitions in the current-voltage domain. The numbers refer to the corresponding curves in Fig. 7.

|  | $\phi_{rst}$ (Vs) | $Q_{rst}$ (µC) | $n$ |
|---|---|---|---|
| Average | 3.28 | 562 | 1.50 |
| Standard dev. | 0.76 | 255 | 0.0999 |

**Tab. 2.** Extracted mean values and deviations for the parameters of the model described in (1).

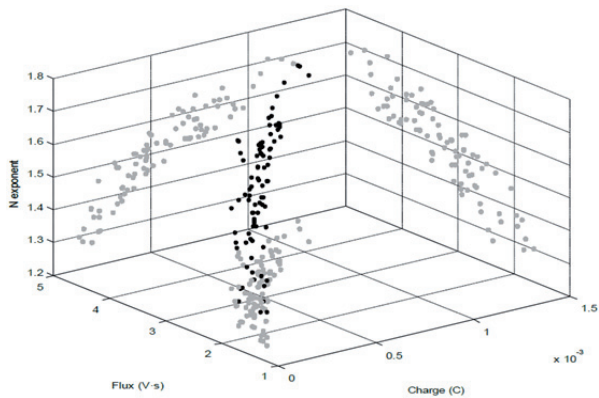

**Fig. 9.** Extracted parameters $\phi_{rst}$, $Q_{rst}$ and $n$ represented in 3D plot. Dark dots are the points in the 3D space, while the clear points correspond to projections into the corresponding axis planes.

In Fig. 9 we have represented the extracted model parameters for each RS cycle considered. From this figure, it can be concluded that there is some correlation between them. In this respect, there might be connections between the values of these parameters and the geometry and number of conductive filaments that control RS processes. We are still deepening on the physics of RS in these devices to shed light on this issue.

## 4. Conclusions

In this work we have presented measurements of 100 reset cycles of a single unipolar RS-based memristor. We have moved from the usual representation in the I-V space to a representation in the Q-φ domain, and we have extracted the reset voltages and reset currents in these new domain. We have seen that, after normalizing all the RS reset cycles with the values of $Q_{rst}$ and $\phi_{rst}$, the curves can be modeled with a single explicit analytical expression in the Q-φ domain, shown in (1). The fitting is fairly good, as shown in Fig. 7 and 8. This simple model has three fitting parameters, but, as can be seen in Fig. 9, they are strongly correlated through some underlying process, which we assume to be the number of filaments and their radii, but this is still under study.

This simple model can be complemented with other effects such as contact effects, as already done in other devices (see [19], for instance). It can be also used to simulate the statistical behavior of resistive switching unipolar memristors.

It must be noted also that it does not depend on the shape of the input waveform, since the model is, as recommended in [4], in the Q-φ domain. New experiments are in process with different waveforms and frequencies to further check the validity of this approach.

## Acknowledgments

We thank Francesca Campabadal and Mireia Bargalló from the IMB-CNM (CSIC) in Barcelona for fabricating and providing the experimental measurements of the devices employed in this manuscript. We also acknowledge the European COST Action MemoCIS IC1401, and the Spanish "Red de Excelencia" Nanovar. Part of this work was funded under the "Programa Pont La Caixa-UIB 2014", and under the Spanish Ministry of Economy and Competitiveness, Projects TEC2013-40677-P, TEC2014-52152-C3-2-R, and TEC2014-56244-R, and under the Junta de Andalucía, project FQM.1861.

## References

[1] XIE, Y. *Emerging Memory Technologies*. Springer, 2014.

[2] WASER, R., AONO, M. Nanoionics-based resistive switching memories. *Nature Materials*, 2007, vol. 6, p. 833–840. DOI: 10.1038/nmat2023

[3] *The International Technology Roadmap for Semiconductors*. 2013 version. [Online] Available at: http://public.itrs.net.

[4] CHUA, L. O. Resistance switching memories are memristors. *Applied Physics A*, 2011, vol. 102, no. 4, p. 765–783. DOI: 10.1007/s00339-011-6264-9

[5] STRUKOV, D. B., SNIDER, G. S., STEWART, D. R., WILLIAMS, R. S. The missing memristor found. *Nature*, 2008, vol. 453, p. 80–83. DOI: 10.1038/nature06932

[6] CHUA, L. O. Memristor - the missing circuit element. *IEEE Transactions on Circuit Theory*, 1971, vol. 18, no. 5, p. 507–519. DOI: 10.1109/TCT.1971.1083337

[7] BIOLEK, Z., BIOLEK, D., BIOLKOVA, V. SPICE model of memristor with nonlinear dopant drift. *Radioengineering*, 2009, vol. 18, no. 2, p. 210–214.

[8] JIMÉNEZ-MOLINOS, F., VILLENA, M. A., ROLDÁN, J. B., ROLDÁN, A. M. A SPICE compact model for unipolar RRAM reset process analysis. *IEEE Transactions on Electron Devices*, 2015, in press.

[9] SHIN, S., KIM, K., KANG, S-M. Compact models for memristors based on charge–flux constitutive relationships. *IEEE Transactions on Computer Aided Design of Integrated Circuits and Systems*, 2010, vol. 29, no. 4, p. 590–598. DOI: 10.1109/TCAD.2010.2042891

[10] PICOS, R., AL CHAWA, M. M., ROCA, M., GARCIA-MORENO, E. A charge-dependent mobility memristor model. In *10th Spanish Conf. on Electron Devices*. Arajuez (Spain), 2015.

[11] VILLENA, M. A., GONZÁLEZ, M. B., JIMÉNEZ-MOLINOS, F., CAMPABADAL, F., ROLDÁN, J. B., SUÑÉ, J., ROMERA, E., MIRANDA, E. Simulation of thermal reset transitions in RRAMs including quantum effects. *Journal of Applied Physics*, 2014, vol. 115, p. 214504. DOI: 10.1063/1.4881500

[12] DEGRAEVE, R., FANTINI, A., RAGHAVAN, N., GOUX, L., CLIMA, S., CHEN, Y. Y., BELMONTE, A., COSEMANS, S., GOVOREANU, B., WOUTERS, D. J., ROUSSEL, P. H., KAR, G. S., GROESENEKEN, G., JURCZAK, M. Hourglass concept for RRAM: a dynamic and statistical device model. In *IEEE 21st International Symposium on the Physical and Failure Analysis of Integrated Circuits (IPFA)*. Singapore, 2014, p. 245–249. DOI: 10.1109/IPFA.2014.6898205

[13] YU, S., GUAN, X., WONG, P. On the stochastic nature of resistive switching in metal oxide RRAM: Physical modeling, Monte

Carlo simulation, and experimental characterization. In *IEEE International Electron Devices Meeting (IEDM)*. Washington (USA), 2011, p. 413–416. DOI: 10.1109/IEDM.2011.6131572

[14] LARENTIS, S., NARDI, F., BALATTI, S., GILMER, D. C., IELMINI, D. Resistive switching by voltage-driven ion migration in bipolar RRAM—Part II: Modeling. *IEEE Transactions on Electron Devices*, 2012, vol. 59, no. 9, p. 2468–2475. DOI: 10.1109/TED.2012.2202320

[15] BOCQUET, M., DELERUYELLE, D., MULLER, C., PORTAL, J.-M. Self-consistent physical modeling of set/reset operations in unipolar resistive-switching memories. *Applied Physics Letters*, 2011, vol. 98, p. 263507. DOI: 10.1063/1.3605591

[16] VILLENA, M. A., JIMÉNEZ-MOLINOS, F., ROLDÁN, J. B., SUÑÉ, J., LONG, S., LIAN, X., GÁMIZ, F., LIU, M. An in-depth simulation study of thermal reset transitions in resistive switching memories. *Journal of Applied Physics,* 2013, vol. 114, p. 144505 to 144505-8. DOI: 10.1063/1.4824292

[17] GONZALEZ, M. B., RAFÍ, J. M., BELDARRAIN, O., ZABALA, M., CAMPABADAL, F. Analysis of the switching variability in Ni/HfO$_2$-based RRAM devices. *IEEE Transactions on Device and Material Reliability*, 2014, vol. 14, no. 2, p. 769–771. DOI: 10.1109/TDMR.2014.2311231

[18] VILLENA, M. A., JIMÉNEZ-MOLINOS, F., ROLDÁN, J. B., SUÑÉ, J., LONG, S., MIRANDA, E., LIU, M. A comprehensive analysis on progressive reset transitions in RRAMs. *Journal of Physics D, Applied Physics*, 2014, vol. 47, p. 205102. DOI: 10.1088/0022-3727/47/20/205102

[19] IÑIGUEZ, B., PICOS, R., VEKSLER, D., KOUDYMOV, A., SHUR, M. S., YTTERDAL, T., JACKSON, W. Universal compact model for long- and short-channel thin-film transistors. *Solid-State Electronics*, 2008, vol. 52, no. 3, p. 400–405. DOI: 10.1016/j.sse.2007.10.027

## About the Authors …

**Rodrigo PICOS** received his degree in Physics and the Ph.D. degree from the Universitat de les Illes Balears (Spain), in 1996 and 2006, respectively. His research interest includes resistive switching memories (RRAMs) and memristors, modeling of organic and small-molecules transistors, and parameter extraction, focusing on the influences of the fabrication process in the parameter dispersion.

**Juan Bautista ROLDÁN** received a degree in Physics and the Ph.D. degree from the Universidad de Granada (Spain), in 1993 and 1997, respectively. His research interest includes simulation and modeling of resistive switching memories (RRAMs) and memristors, modeling of nanometric conventional and multigate devices mainly based on SOI technology and Verilog-A implementation of device models for circuit simulation.

**Mohamad Moner AL CHAWA** was born in Damascus, Syria in 1984. He received his B.S. degree in Electrical Engineering, Electronics & Communication from the International University for Science & Technology, Daraa (Syria), and the M.S. degree in Advanced Communication Engineering from Damascus University (Syria), in 2010 and 2013 respectively. Currently, he is working toward his Ph.D. degree in Electronic Engineering at the Universitat de les Illes Balears (Spain).

**Pedro GARCÍA-FERNÁNDEZ** received a degree in Electronics Engineering and the Ph.D. degree from the Universidad de Granada (Spain), in 1997 and 2000, respectively. His research interest includes simulation and modeling of resistive switching memories (RRAMs) and memristors. He has also interest in the design of circuits for memory applications.

**Francisco JIMÉNEZ-MOLINOS** received a degree in Physics and a degree in Electronic Engineering in 1998 and 2000, respectively, and the Ph.D. degree in 2002, all from the University of Granada (Spain). Since 1999, he has been working on the MOS device physics. His current research interest includes the simulation and modeling of RRAMs.

**Eugeni GARCÍA-MORENO** received the M.S. degree from the Polytechnic University of Catalonia, Barcelona, Spain, in 1975 and the Ph.D. degree from University Paul Sabatier, Toulouse, France, in 1979. He is currently a Professor with the Physics Dept., University of the Balearic Islands (UIB), Palma, Spain, where he is the Head of the Electronic Engineering Group. His research interests include compact device models, analog integrated circuit test techniques, and CMOS radiation sensors.